\documentclass[aps,prb,reprint,showpacs,showkeys]{revtex4-2}
\usepackage{graphicx} 
\usepackage{amsmath} 
\usepackage[version=4]{mhchem} 
\usepackage{bm} 
\usepackage{natbib}
\usepackage{natmove}
\usepackage[colorlinks=true, linkcolor=blue, citecolor=blue, urlcolor=blue, linktoc=page, bookmarks=false, pdfstartview={FitH}, pdfborder={0 0 0.0 [3 3]}]{hyperref} 
\usepackage{color} 
\usepackage{dcolumn} 
\newcolumntype{.}{D{.}{.}{3.2}}
\newcolumntype{-}{D{.}{.}{4.2}}
\usepackage{makecell}
\usepackage{multirow}
\usepackage{cleveref}
\usepackage{easyReview} 
\crefname{figure}{Fig.}{Figs}
\crefname{table}{Table}{Tables}
\renewcommand{\today}{\number\day \space \ifcase \month \or January\or February\or March\or April\or May\or June\or July\or August\or September\or October\or November\or December\fi \space \number\year} 
\def\m1r{\multicolumn{1}{r}}
\newcolumntype{P}[1]{>{\centering\arraybackslash}p{#1}}
\begin{document}
\title{Unsupervised Deep Neural Network Approach To Solve Fermionic Systems}
\author{Avishek \surname{Singh}}
\email[Email: ]{avishek@iiserb.ac.in}
\author{Nirmal \surname{Ganguli}}
\email[Email: ]{NGanguli@iiserb.ac.in}
\affiliation{Department of Physics, Indian Institute of Science Education and Research Bhopal, Bhauri, Bhopal 462066, India}
\date{\today}
\begin{abstract}
Solving the Schr\"{o}dinger equation for interacting many-body quantum systems faces computational challenges due to exponential scaling with system size. This complexity limits the study of important phenomena in materials science and physics. We develop an Artificial Neural Network (ANN)-driven algorithm to simulate fermionic systems on lattices. Our method uses Pauli matrices to represent quantum states, incorporates Markov Chain Monte Carlo sampling, and leverages an adaptive momentum optimizer. We demonstrate the algorithm's accuracy by simulating the Heisenberg Hamiltonian on a one-dimensional lattice, achieving results with an error in the order of $10^{-4}$ compared to exact diagonalization. Furthermore, we successfully model a magnetic phase transition in a two-dimensional lattice under an applied magnetic field.  Importantly, our approach avoids the sign problem common to traditional Fermionic Monte Carlo methods, enabling the investigation of frustrated systems. This work demonstrates the potential of ANN-based algorithms for efficient simulation of complex quantum systems, opening avenues for discoveries in condensed matter physics and materials science.
\end{abstract}
\maketitle
\section{\label{sec:intro}Introduction}
The Schr\"{o}dinger equation governs quantum systems, underpinning crucial physics phenomena. Yet, solving it for systems with many interacting particles is NP-hard due to the exponential growth of degrees of freedom in the wave function. Modern physics aims to mitigate this complexity by reducing the dimensionality of many-body systems. Efforts involve developing efficient algorithms and leveraging increased computing power. Despite progress, traditional methods face challenges in handling systems with numerous interacting particles due to the complexity of the wave function. Addressing these challenges remains a focal point in modern physics.

Over the past two decades, quantum computing and machine learning advancements have revolutionized the study of complex physical systems across multiple scales. These approaches offer efficient modeling methods for quantum systems, vital in condensed matter physics and quantum chemistry. A notable recent breakthrough is the utilization of neural-network quantum states for investigating many-body systems, with promising outcomes across various physics and chemistry problems \cite{Carleo2017,Torlai2017,David2020,David2014,Carleo2019,Amber2022,Corey2021,Lovato2022}. While these methods closely align with analytical solutions and quantum Monte Carlo calculations, challenges persist, particularly in handling strongly correlated or entangled many-body systems. Nonetheless, the potential of these techniques has sparked considerable excitement and investment in the scientific community, driving further refinement and integration with other quantum simulation approaches. Emerging applications include material development and quantum technology design. Despite challenges, researchers are committed to exploring the capabilities of these methods to advance quantum science and technology boundaries.

The pursuit of understanding quantum mechanical systems' fundamental properties and behavior has long captivated physicists and researchers. The key to this endeavour is determining the ground-state properties of systems, which offers invaluable insights. Varied techniques such as variational and diffusion Monte Carlo algorithms \cite{MISAWA2019447, Pang2014} have emerged as pivotal methods for this purpose. These algorithms rely on a trial wave function, which approximates the system's true ground-state wave function, defined with variational parameters optimized to minimize total energy. However, defining an accurate and flexible trial wave function proves challenging, often requiring prior knowledge and physical intuition about the system. Without such information, researchers explore methods, including simple analytical expressions based on known system properties or machine learning techniques like neural networks to learn wave functions from data or simulations. Despite these advances, challenges persist, particularly for systems with strong correlations or entanglement, where wave function approximation remains complex. Nonetheless, Monte Carlo algorithms and computational methods remain indispensable for studying quantum systems, driving further exploration for defining accurate and flexible trial wave functions \cite{HORNIK1989359, Murphy2012, Goodfellow2017, Hastie2001, bishop2006pattern}.

The universal approximation theorem asserts a deep neural network's capability to approximate any continuous function within a certain error margin \cite{HORNIK1989359}. This theorem has spurred interest in employing deep learning methods to tackle complex physical problems, including the study of quantum mechanical ground states using variational and diffusion Monte Carlo algorithms \cite{MISAWA2019447, Pang2014}. Researchers have increasingly turned to neural networks to represent variational state wave functions, overcoming the limitations of traditional approaches. This approach has shown promise and has been successfully implemented across various studies \cite{tomoki2016, Carrasquilla2017, VanNieuwenburg2017, Zhang2017, Tomi2017, Broecker2017, Tanaka2017, Broecker2017-1, Kelvin2018, Zhang2018, Mano2017}. Notably, neural networks derived from restricted Boltzmann machines have been instrumental \cite{Carleo2017, Carleo2019}. However, accurately incorporating problem symmetries remains critical for achieving precise results, necessitating attention to ensure neural networks effectively capture the system's symmetries. 

This work presents a method developed using a deep neural network and Markov Chain Monte Carlo, leveraging the fundamental principle of quantum mechanics (variational principle) to solve fermionic systems. We showed that a fully connected feed-forward neural network could solve the complex fermionic system parameters and construct the full quantum state. We demonstrated that the fully connected feed-forward neural network with one input, one hidden, and one output layer could achieve the ground state of the Heisenberg model on a one-dimensional and two-dimensional lattice. We have used the adaptive momentum optimizer to optimise the neural network. We demonstrated that the neural network-driven many-body simulation could understand the applied external factors on the fermionic system and compute the properties by considering them. Also, the presented method is robust and capable of simulating any shape or size of the lattice, which means it can also simulate the frustrated lattices, for which most traditional methods like Quantum Monte Carlo fail.

\section{\label{sec:method}Method of calculations}
We will construct the method using a quantum system, neural network architecture, and neural network optimization.

\subsection{\label{subsec:Method-QuantumSystem} The Quantum System}
\begin{figure}
    \centering
    \includegraphics[scale = 0.34]{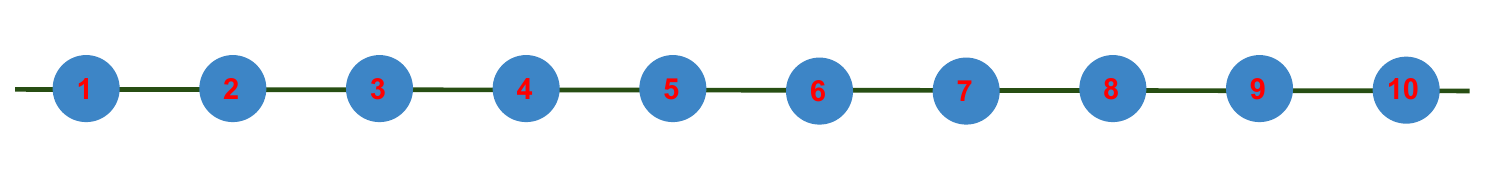}
    \caption{One-dimensional lattice with $M = 10$ sites to simulate fermionic quantum system.}
    \label{fig:Method-QuantumSystem-Lattice1D}
\end{figure}

\begin{figure}
    \centering
    \includegraphics[scale = 0.34]{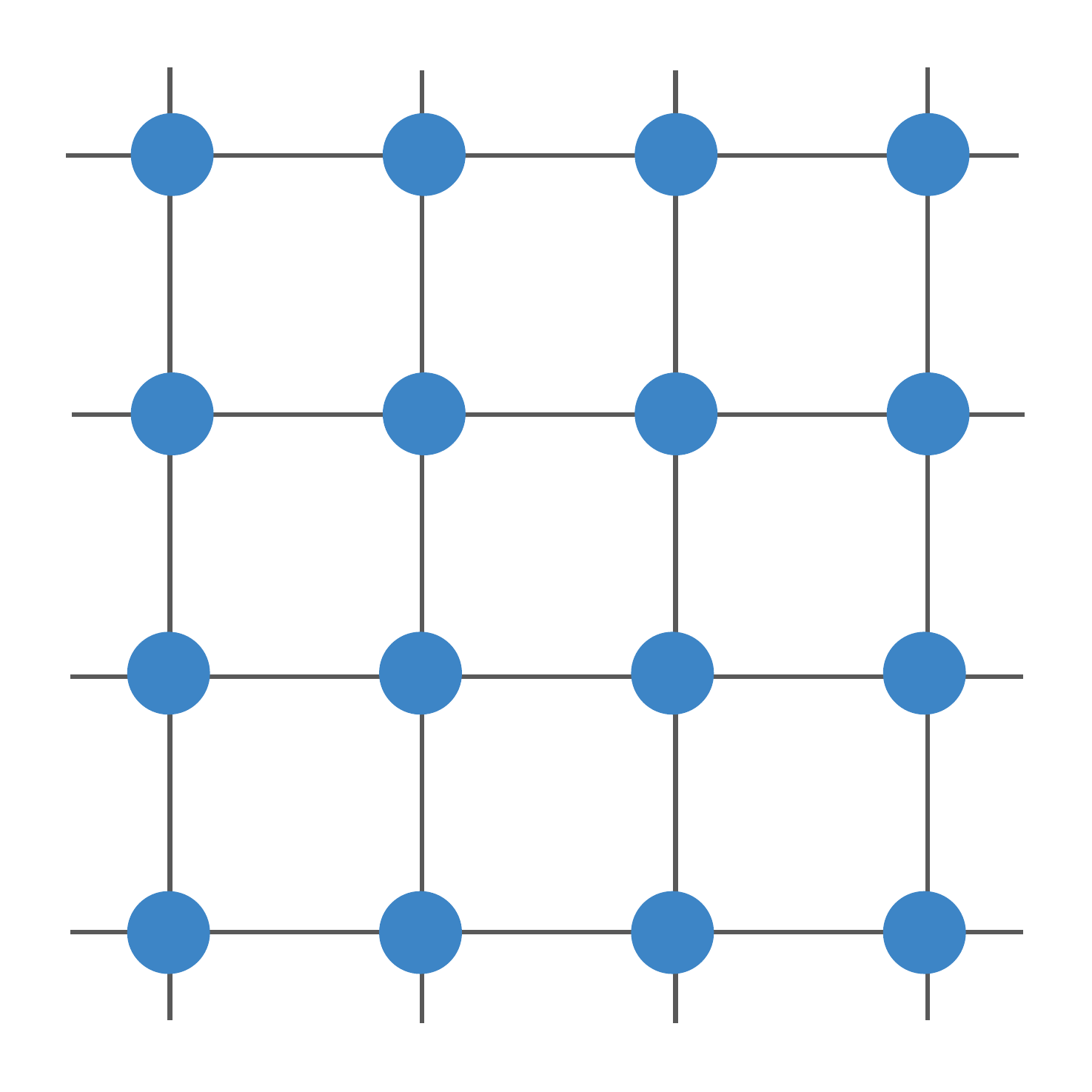}
    \caption{Two-dimensional lattice with $M = 16$ sites ($4 \times 4$) to simulate fermionic quantum system.}
    \label{fig:Method-QuantumSystem-Lattice2D}
\end{figure}

Before describing the many-body quantum states, first, we will define the lattice structure. In this work, we will simulate the fermionic quantum system for a given Hamiltonian ($H$) on a one-dimensional lattice with $M = 10$ sites, depicted in \cref{fig:Method-QuantumSystem-Lattice1D}, occupied by $N = 10$ fermions and on a two-dimensional lattice with $M = 16$ sites ($4 \times 4$), depicted in \cref{fig:Method-QuantumSystem-Lattice2D}, occupied by $N = 16$ fermions arranged in an antiferromagnetic ordering. $n_i$ represents the number of fermions per site, satisfying the condition $N = \sum_i n_i$ and $0 \leq n_i \leq 2$ to satisfy Pauli's exclusion principle.

The many-body quantum state for the $N$ fermions occupying the lattice with $M$ sites is given as,
\begin{equation}
    \label{eq:Method-QuantumSystem-State}
    | \Psi \rangle = \sum_{i = 1}^{M } \phi(n_1, n_2, ..., n_{M}) | n_1, n_2, ..., n_{M} \rangle
\end{equation}

For fermions many-body quantum state can be represented as follows,
\begin{equation}
    \label{eq:Method-QuantumSystem-Qstate}
    | \Psi \rangle = \sum_{s_i} \phi(\uparrow_1, \downarrow_2, \uparrow_3, .... , \downarrow_M) | \uparrow_1, \downarrow_2, \uparrow_3, .... , \downarrow_M \rangle
\end{equation}
where $\sum_{s_i}$ is the summation over spins, and $\uparrow$ and $\downarrow$ are represented as spin-up and spin-down, respectively, to compute the expectation value.

In Pauli matrices representation, the quantum state is represented as,
\begin{equation}
    \label{eq:Method-QuantumSystem-QstatePauli}
    | \Psi \rangle = \sum_{s_i} \phi(\bm{\sigma}) | \sigma_1, \sigma_2, \sigma_3, .... , \sigma_M \rangle
\end{equation}
where, $\sigma_i = \sigma_i^x \hat{X} + \sigma_i^y \hat{Y} + \sigma_i^z \hat{Z}$. 

However, we have given the full form of the Pauli matrices for the general case here. Nevertheless, in our simulation, we will localize the spins in ($\pm$)ve z-axis, so here $\sigma_i \approx \sigma_i^z$, which means other components of Pauli matrices are zero. Note that the algorithm is not limited to this approximation; one can use the full form of the Pauli matrices and can localize spins in any direction or angle. The complex number $\phi(\bm{\sigma})$ is computed using the output of the neural network depicted in \cref{fig:Method-NeuralNet-ANN}.

Now that we have defined the quantum state, we will define the Hamiltonian for the calculation. In our simulation, we used two Hamiltonians. The first one is the Heisenberg Hamiltonian, given as follows,
\begin{equation}
    \label{eq:Method-QuantumSystem-HeisenberHamiltonian}
    H = -J \sum_{\langle i,j \rangle} \Vec{S_i^z} \cdot \Vec{S_j^z}
\end{equation}
where $J$ is the exchange interaction term. While $S_i^z$ is spin at $i^{th}$ site of lattice localized in the z-axis.

The second Hamiltonian is the Heisenberg Hamiltonian in an external magnetic field, given as
\begin{equation}
    \label{eq:Method-QuantumSystem-HeisenberHamiltonianExMag}
     H = -J \sum_{\langle i,j \rangle} \Vec{S_i^z} \cdot \Vec{S_j^z} - g \mu_B h \sum_i  \Vec{S_i^z}
\end{equation}
Where $J$ is the exchange interaction term, $g$ is Landé g-factor, $\mu_B$ is Bohr magnetic moment, and $h$ is the magnetic field strength. For our calculations, we have taken $g \mu_B = 1$. 

After defining the Hamiltonian, we will define the expectation value of Hamiltonian to compute the ground state energy of the quantum system. We have defined the Hamiltonian as spins $S_i^z$, but the quantum state $| \Psi \rangle$ is described as Pauli matrices $\sigma$. We must define the Hamiltonian as Pauli matrices $\sigma$ to compute the expectation value. This can be easily achieved by changing the spin $S_i^z$ to Pauli matrices $\sigma$ using formulation $S_i^z = \frac{1}{2} \hbar \sigma_i^z$, and for the simulation, we will put $\hbar = 1$.

The expectation value of Hamiltonian is calculated as,
\begin{align}
    \label{eq:Method-QuantumSystem-Expect}
    \langle \Tilde{H} \rangle =  \left< \sum_{\bm{\sigma'}} \langle \bm{\sigma} | H | \bm{\sigma'} \rangle \frac{\phi(\bm{\sigma'})}{\phi(\bm{\sigma})} \right>_{N_T}
\end{align}
where, $\phi(\bm{\sigma'})$ and $\phi(\bm{\sigma})$ are calculated suing the output of the neural network, $N_T$ is the total number of possible states.

\subsection{\label{subsec:Method-NeuralNet} Neural Network Architecture}
\begin{figure}
    \centering
    \includegraphics[scale = 1]{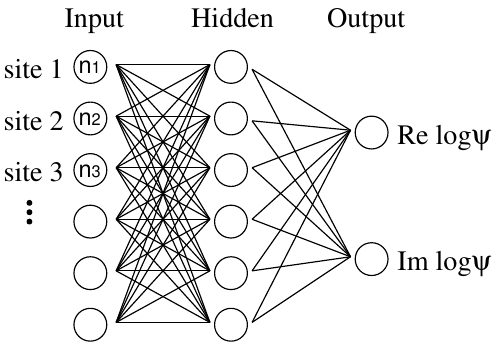}
    \caption{The neural network architecture schematic to simulate a fermionic quantum system.}
    \label{fig:Method-NeuralNet-ANN}
\end{figure}

\cref{fig:Method-NeuralNet-ANN} depicts the deep neural network architecture used to simulate the fermionic quantum system in the lattices depicted in \cref{fig:Method-QuantumSystem-Lattice1D} and \cref{fig:Method-QuantumSystem-Lattice2D} with one input, one hidden, and one output layer. The input layer has $M \in [10, 16]$ neurons as we have $M \in [10, 16]$ sites in the lattice, the hidden layer has $2M + 1$ neurons, and the output layer has $2$ neurons. Here, we have used $2M + 1$ as a general convention to set the number of neurons in the hidden layer, which is identified by evaluating the neural network for different combinations of hidden layer neurons. Note that if the number of the lattice site (number of neurons in the input layer) changes significantly compared to our settings, the convention of one hidden layer with $2M + 1$ neurons might not be valid. Moreover, one has to identify the correct combination for the number of hidden layers and the number of neurons in each hidden layer.

The value of each neuron in the input layer is given as,
\begin{equation}
    \label{eq:Method-NeuralNet-Input}
    Y_i^{(0)} = | n_i \rangle
\end{equation}
Note that we have used number representation $n_i$ as input to the neural network. We will use the Pauli Matrices formalism to compute the expectation value of the Hamiltonian. This arrangement is made because we can not directly input Pauli Matrices into the neural network. Instead, we input the number of fermions per lattice site $n_i$ to compute the $\phi(\bm{\sigma})$.

The value of each neuron in the hidden layer is calculated as,
\begin{equation}
    \label{eq:Method-NeuralNet-Hidden}
    Y_j^{(1)} = \sum_{i=1}^{M} W_{ij} Y_i^{(0)} + b_j
\end{equation}
where $M$ is the number of neurons in the input layer, which is equal to the number of sites in the lattice, $i$ is the index over the input layer, $j$ is the index over the hidden layer, $W_{ij}$ are the weights between the input and hidden layer, and $b_j$ are the bias factors for the hidden layer.

The value of each neuron in the output layer is calculated as,
\begin{equation}
    \label{eq:Method-NeuralNet-Optput}
    Y_k^{(2)} = \sum_{j=1}^{2M + 1} W_{jk} f(Y_j^{(1)})
\end{equation}
where $k$ is the index over the output layer, $W_{jk}$ are the weights between the output and hidden layer, and $f(x)$ is the nonlinear function of the hidden layer, which is given as $f(x) = \tanh(x)$.

The complex coefficient $\phi(\bm{\sigma})$ in \cref{eq:Method-QuantumSystem-QstatePauli} is calculated using the values of the output layer of the neural network given in \cref{eq:Method-NeuralNet-Optput} as
\begin{equation}
    \label{eq:Method-QuantumSystem-Phi}
    \phi(\bm{\sigma}) = \exp \left( Y_1^{(2)} \right) + i Y_2^{(2)}
\end{equation}

\subsection{\label{subsec:Method-Optimization} Neural Network Optimization}
\cref{eq:Method-QuantumSystem-Expect} will be used as a cost function to optimize the neural network model presented in \cref{fig:Method-NeuralNet-ANN}. Before optimization of the neural network, we need to set up some rules to sample the occupation states from Hilbert space.

To sample the occupation states from Hilbert space, we have used the ``Exchange Sampler" to keep the antiferromagnetic ordering consistent throughout the simulation. The \textit{Exchange Sampler} works on two local degrees of freedom, which means it will change particle occupation on two lattice sites to construct the new occupation state. The rules are as follows for the occupation state defined as $|\bm{\uparrow, \downarrow}\rangle = | \uparrow_1, \downarrow2, \uparrow_3, \ldots, \downarrow_i, \ldots, \uparrow_j, \ldots, \downarrow_M \rangle$,
\begin{enumerate}
    \item This sampler acts locally on two degrees of freedom $\downarrow_i$ and $\uparrow_j$ and proposes a new state as follows,
            \begin{equation}
                \label{eq:Method-Optimization-OcuupationState_Exchange}
                |\bm{\uparrow', \downarrow'}\rangle = | \uparrow_1, \downarrow_2, \uparrow_3, \ldots, \uparrow'_i, \ldots, \downarrow'_j, \ldots, \downarrow_M \rangle
            \end{equation}
            where $\uparrow'_i \neq \downarrow_i$ and $\downarrow'_j \neq \uparrow_j$
    \item The transition probability associated with this sampler can be decomposed into two steps. First is a pair of indices $(i, j) \in \{1, \ldots, M\}$, and such that $\text{dist}(i,j \leq d_\text{max})$ is chosen with uniform probability and the second is among all the possible $\uparrow'_i$, $\downarrow'_j$ can take, one of them is chosen with uniform probability.
    \item We choose two sites $(i, j) \in \{1, \ldots, M\}$ with uniform probability, and then we flip both of the fermions on the chosen sites, \textit{i.e.} $\uparrow'_i = \uparrow_j$ and $\downarrow'_j = \downarrow_i$.
    \item We can not exchange the fermions for two states if they are in opposite directions and only one of the states occupies two fermions. For Example, If sate-1 is $|\uparrow \downarrow \rangle$ and state-2 is $|\cdot \uparrow \rangle$, then we can not exchange the last spins of both the states.
    \item We can exchange the fermions for two states if one of them does not have any fermion in it. From the example given in point-4, we can exchange the second spin of state-1 with the first spin in state-2. However, here also, the rule in point-4 applies; if the exchange makes both spins aligned in the same direction in either of the states, then the exchange is not allowed.
\end{enumerate}

Though the sampler discussed above proposes a new occupation state $|\bm{\uparrow', \downarrow'}\rangle$ from the initial occupation state $|\bm{\uparrow, \downarrow}\rangle$ using the rules discussed above, to decide whether the newly proposed state should be accepted or rejected is chosen based on the probability given by probability distribution. 

The probability of accepting the newly proposed state $|\bm{\uparrow', \downarrow'}\rangle$ from initial state $|\bm{\uparrow, \downarrow}\rangle$ is give as,
\begin{equation}
    \label{eq:Method-Optimization-AcceptanceProb}
    P_{(\bm{\sigma} \to \bm{\sigma'})} = \text{min} \left[ 1, \left| \frac{\phi(\bm{\sigma'})}{\phi(\bm{\sigma})} \right|^2 \right]
\end{equation}

Here note that the squared probability $\left| \frac{\phi(\bm{\sigma'})}{\phi(\bm{\sigma})} \right|^2$ allows us to simulate even frustrated lattices by avoiding sign problem. And $\phi(\bm{\sigma})$ is given by \cref{eq:Method-QuantumSystem-Phi}.

The probability distribution for occupation state $|\bm{\sigma}\rangle$ is given as,
\begin{equation}
    \label{eq:Method-Optimization-ProbDist}
    p(\bm{\sigma}) = \frac{\left| \phi(\bm{\sigma}) \right|^2}{\sum_{\bm{\sigma'}} \left| \phi(\bm{\sigma'}) \right|^2}
\end{equation}

After defining the ``Exchange Sampler" as a sampling algorithm, we will formulate the optimization algorithm using one of the optimizers. For our case, we will use ``Adaptive Momentum (Adm)" optimizer to train the neural network formulated in \cref{subsec:Method-NeuralNet} and schematically presented in \cref{fig:Method-NeuralNet-ANN}. 

The ``Adaptive Momentum (Adm)" optimizer is formulated as,
\begin{align}
    \label{eq:Method-Optimization-Adm}
    w_i & \rightarrow w_i - \gamma \frac{m_i}{1-\beta^l_1}\frac{1}{\sqrt{\frac{v_i}{i-\beta^l_2}}+\epsilon} \nonumber \\
    m_i & \rightarrow \beta_1 m_i + (1-\beta_1)\frac{\partial \langle H \rangle}{\partial w_i} \\
    v_i & \rightarrow \beta_2 v_i + (1-\beta_2) \left(\frac{\partial \langle H \rangle}{\partial w_i}\right)^2 \nonumber
\end{align}

where, $\gamma < 1$ , $\beta_1 = 0.9$ and $\beta_2 = 0.999$ for $l^{th}$ update. The initial values of $v_i$ \& $m_i$ are zero. $w_i$ are the neural network parameters.

\section{\label{sec:result}Results and Discussions}
This section presents the results to validate the artificial intelligence-driven quantum many-body algorithm for the fermionic many-body quantum system obtained using the methodology presented in \cref{sec:method}. Here, we will show the result of two simulations done for fermionic systems. The first is Heisenberg Hamiltonian, given by \cref{eq:Method-QuantumSystem-HeisenberHamiltonian}, on a one-dimensional lattice chain with $M = 10$ sites occupied by $N = 10$ fermions in antiferromagnetic ordering. The second is Heisenberg Hamiltonian in an external magnetic field, given by \cref{eq:Method-QuantumSystem-HeisenberHamiltonianExMag}, on a two-dimensional lattice ($4 \times 4$) with $M = 16$ sites occupied by $N = 16$ fermions in antiferromagnetic ordering.

\subsection{\label{subsec:Results-HeisenberHamiltonian} The Heisenberg Hamiltonian On A One-dimensional Lattice Chain}
Here, we present the results for the Heisenberg Hamiltonian on a one-dimensional lattice chain. These results are obtained using the method discussed in \cref{sec:method}. The simulation was performed for $300$ iterations considering $\gamma = 0.9$, $\beta_1 = 0.9$, and $\beta_2 = 0.99$ as hyperparameters for adaptive momentum optimizer. In Hamiltonian \cref{eq:Method-QuantumSystem-HeisenberHamiltonian}, $J$ value is taken to be $1.0$.

\begin{figure}
    \centering
    \includegraphics[scale = 0.54]{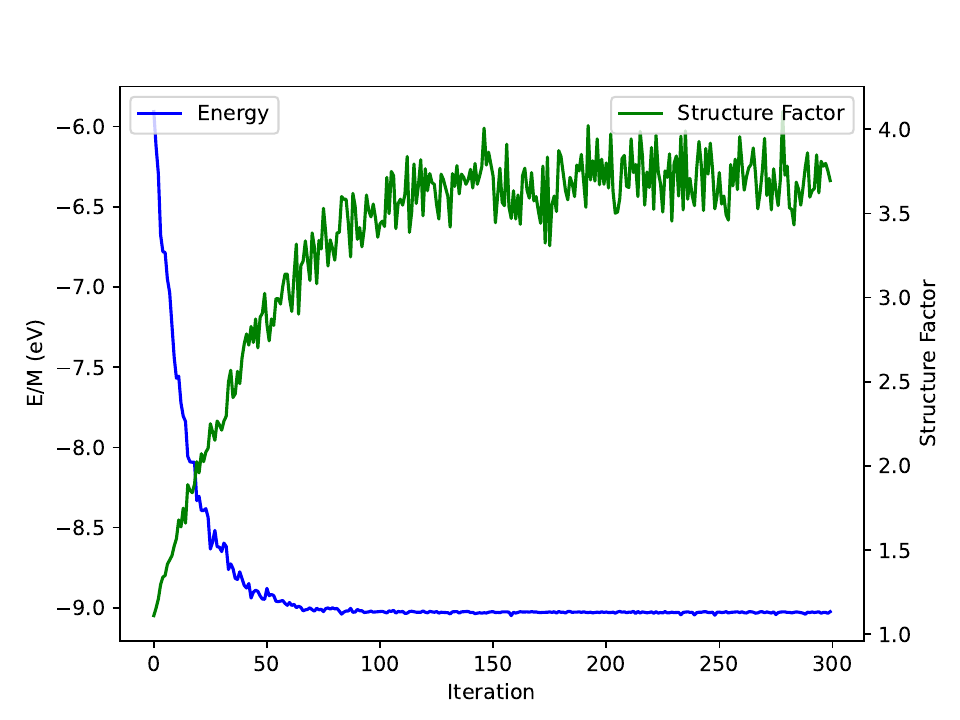}
    \caption{Iterations vs Energy(blue) and Structure Factor(green) plot for the Heisenberg Hamiltonian on a One-dimensional lattice chain with $M = 10$ sites obtained using a fully connected feed-forward neural network.}
    \label{fig:Results-HeisenberHamiltonian-Energy1DHeisen}
\end{figure}
The \cref{fig:Results-HeisenberHamiltonian-Energy1DHeisen} shows the graph between the total energy of the system and the iterations. From the plot, we can see that the system's total energy starts at $\sim -5.5$~eV and converges to $\sim -9.0$~eV, which is the ground state energy of the system. This indicates that the many-body quantum state has reached the proximity of the exact ground state of the system, and the neural network has been optimized properly. The per-site energy of the system is $\sim -0.9028$~eV. To validate the results of per-site and total energy of the system, we performed the simulation using Lanczos exact diagonalization method\cite{Zhang_2010}. The per-site energy of the system using the exact diagonalization method is $\sim -0.9031$~eV. This shows that our algorithm has accuracy in the $\mathcal{O} (10^{-4})$ as compared to the exact diagonalization method. With this, we can conclude that the algorithm is very accurate and can easily solve many-body quantum systems. The \cref{fig:Results-HeisenberHamiltonian-Energy1DHeisen} also shows the graph between the antiferromagnetic structure factor and the iterations. The antiferromagnetic structure factor is a measure of how well an antiferromagnetic material’s magnetic moments are aligned, which is given as,
\begin{equation}
    \label{eq:Results-HeisenberHamiltonian-SF}
    S = \frac{1}{N} \sum_{\langle i,j \rangle} \langle \sigma_i^z \cdot \sigma_j^z \rangle e^{i \pi (i-j)}
\end{equation}
where $N$ is the total number of fermions. From our simulation, we found that the antiferromagnetic structure factor of the One-dimensional lattice chain is equal to $\sim 3.72$.

After simulating the Heisenberg Hamiltonian on a one-dimensional lattice chain with $M = 10$ sites occupied by $N = 10$ fermions, we found that the neural network was adequately optimised, and the neural network parameters are adjusted to store the information of many-body quantum states using adaptive momentum optimizer. The total and per-site energy of the system is calculated with very high accuracy compared to exact diagonalization.

\subsection{\label{subsec:Results-HeisenberHamiltonianExMag} The Heisenberg Hamiltonian In External Magnetic Field On A Two-dimensional Lattice}
After simulating the Heisenberg Hamiltonian on a one-dimensional lattice chain, we applied external magnetic field $\Vec{h} = h \hat{z}$ in the z-direction to see how the algorithm would handle the external modification in the simulation. The Heisenberg Hamiltonian with the external magnetic field is presented in \cref{eq:Method-QuantumSystem-HeisenberHamiltonianExMag}. In this calculation, we have used a two-dimensional $4 \times 4$ lattice with $M = 16$ sites occupied by $N = 16$ fermions. The simulation was performed for $300$ iterations considering $\gamma = 0.9$, $\beta_1 = 0.9$, and $\beta_2 = 0.99$ as hyperparameters for adaptive momentum optimizer. In \cref{eq:Method-QuantumSystem-HeisenberHamiltonianExMag}, $J$ is taken to be $\{-2.0, -1.5, -1.0\}$ and $h$ is varied from $0.0 \mu_B$ to $6.0 \mu_B$.
\begin{figure}
    \centering
    \includegraphics[scale = 0.35]{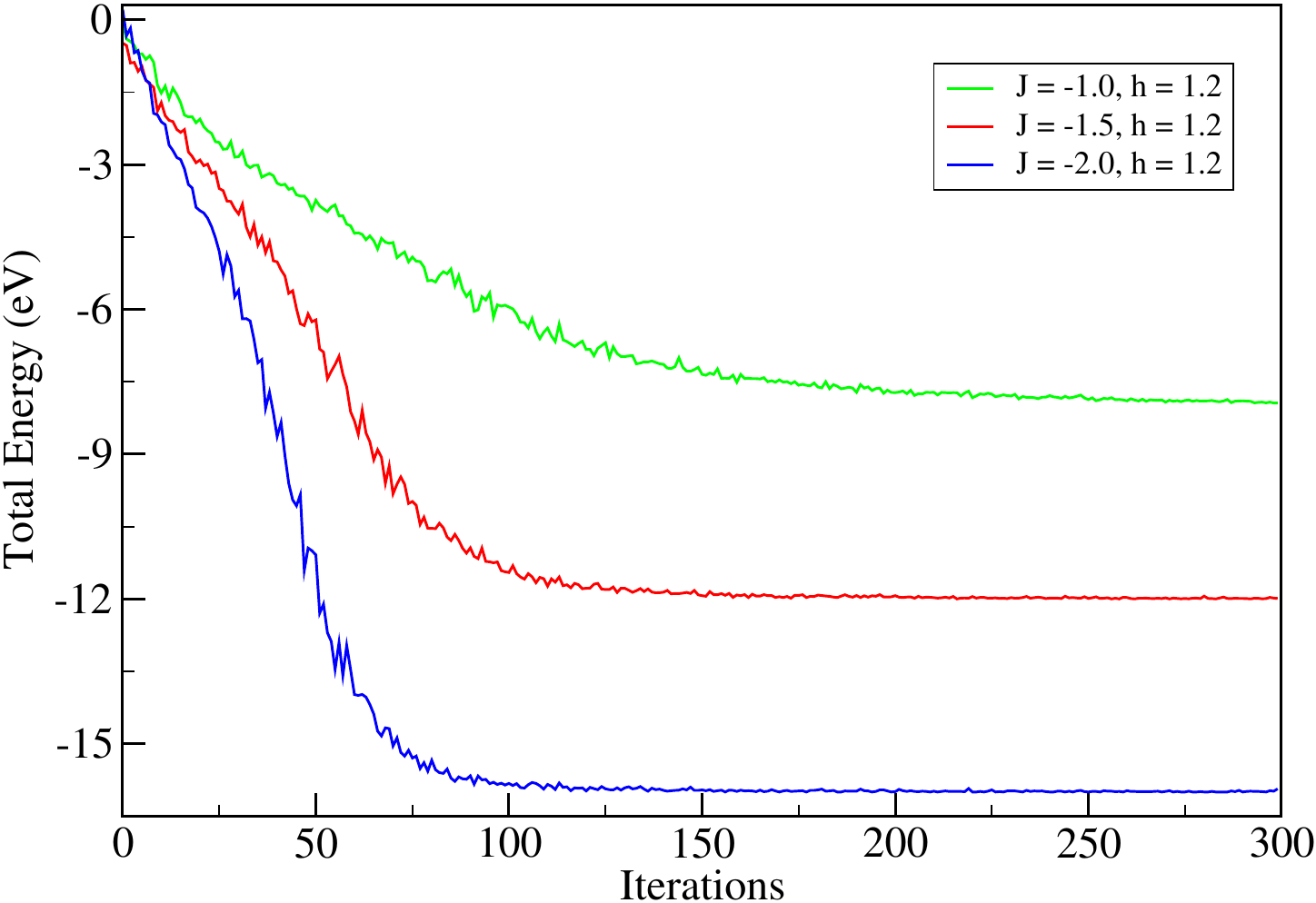}
    \caption{The total energy vs iterations for $J \in [-1.0, -1.5, -2.0]$ and $h = 1.2 \mu_B$ for simulation done for the Hamiltonian given by \cref{eq:Method-QuantumSystem-HeisenberHamiltonianExMag} on the lattice depicted in \cref{fig:Method-QuantumSystem-Lattice2D}.}
    \label{fig:Results-HeisenberHamiltonianExMag-Energy2DHeisenExM}
\end{figure}
\begin{table}
    \caption{\label{tab:Results-HeisenberHamiltonianExMag-EnergyMSZ}The total energy $\varepsilon$~(eV) and net magnetization $M_{s^z}$, for $J \in [-1.0, -1.5, -2.0]$ with increasing magnetic field $h (\mu_B)$, of the fermionic system simulation for Hamiltonian, given by \cref{eq:Method-QuantumSystem-HeisenberHamiltonianExMag}, on the lattice depicted in \cref{fig:Method-QuantumSystem-Lattice2D}.}
    \centering
    \begin{tabular}{.|---|...}
    \hline\hline
        h & \multicolumn{3}{c|}{Total Energy ($\varepsilon$) (eV)} & \multicolumn{3}{c}{Net Magnetization ($\mu_B$)} \\
        \cline{2-7}
         & J = -2.0 & -1.5 & -1.0 & -2.0 & -1.5 & -1.0 \\
        \hline
        0.0 & -15.99 & -11.99 & -7.98 & 0.00 & 0.00 & 0.00 \\
        0.5 & -15.99 & -11.98 & -7.98 & 0.00 & 0.00 & 0.00 \\
        1.0 & -15.99 & -11.98 & -7.96 & 0.00 & 0.00 & 0.03 \\
        1.5 & -15.99 & -11.98 & -7.77 & 0.00 & 0.00 & 0.44 \\
        2.0 & -15.98 & -11.96 & -7.87 & 0.00 & 0.03 & 4.11 \\
        2.5 & -15.98 & -11.79 & -11.80 & 0.00 & 0.43 & 7.31 \\
        3.0 & -15.97 & -11.97 & -15.94 & 0.02 & 4.04 & 7.94 \\
        3.5 & -15.82 & -15.37 & -19.81 & 0.32 & 7.03 & 7.98 \\
        4.0 & -16.07 & -20.09 & -24.15 & 4.40 & 7.93 & 7.99 \\
        4.5 & -19.50 & -23.98 & -27.99 & 7.01 & 7.98 & 7.99 \\
        5.0 & -23.75 & -27.83 & -31.83 & 7.92 & 7.99 & 7.99 \\
        5.5 & -28.14 & -32.15 & -36.15 & 7.99 & 7.99 & 7.99 \\
        6.0 & -31.99 & -35.99 & -39.99 & 7.99 & 7.99 & 7.99 \\
    \hline
    \hline
    \end{tabular}
\end{table}

The total energy of the system vs iteration plot in \cref{fig:Results-HeisenberHamiltonianExMag-Energy2DHeisenExM} shows the convergence of the expectation value of the Hamiltonian with respect to $300$ iterations for $J \in [-1.0, -1.5, -2.0]$ and $h = 1.2 \mu_B$. The plot shows that the total energy of the system starts from $\sim 0$~eV and eventually converges to $\sim -7.91, -11.99, \text{and} -15.99$~eV for $J \in [-1.0, -1.5, -2.0]$, respectively. This indicates that the many-body quantum state has reached the proximity of the exact ground state of the system, and the neural network has been appropriately optimised using the adaptive momentum optimizer. The plot shows that the optimization was stable, as we did not see any wicks or choppiness in the curve. This means the optimizer is working very well.

\begin{figure}
    \centering
    \includegraphics[scale = 0.58]{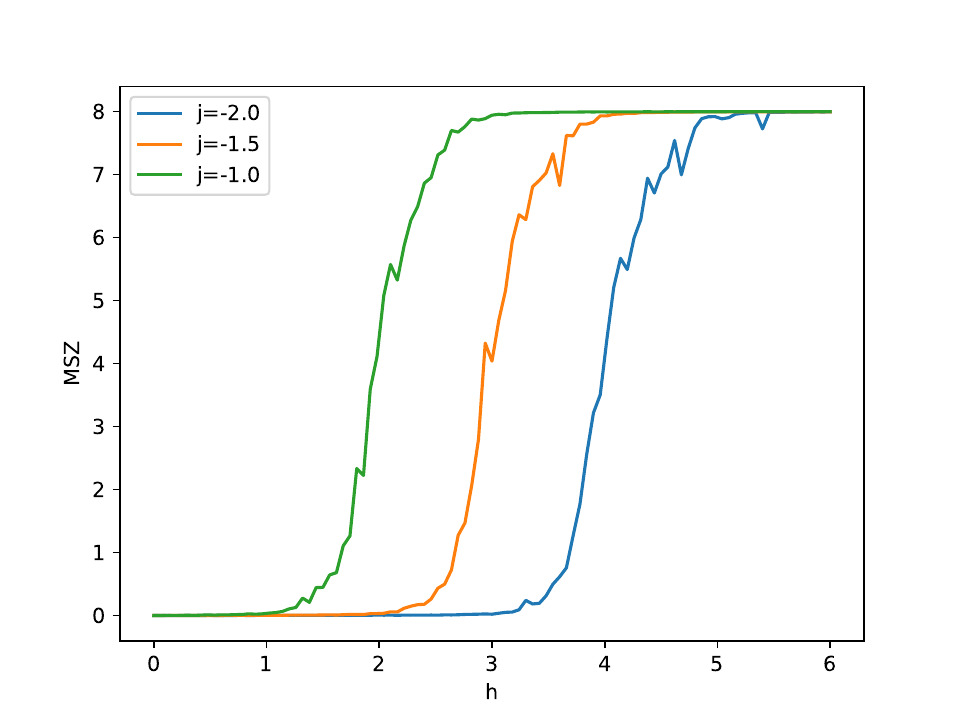}
    \caption{The net magnetization vs external magnetic field graph for $J \in [-1.0, -1.5, -2.0]$ with increasing magnetic field $h (\mu_B)$.}
    \label{fig:Results-HeisenberHamiltonianExMag-MSZ}
\end{figure}

We also studied the effect of the external magnetic field on the system by computing the net magnetization of the system. Theoretically, the net magnetization of any antiferromagnetic system should be zero, which means the spins on the lattice are aligned antiparallel. If the net magnetization is not zero, the spins are not aligned exactly antiparallel. The net magnetization is given as

\begin{equation}
    \label{eq:Results-HeisenberHamiltonianExMag-NetMag}
    M_{S_Z} = \sum_i S_i^z,
\end{equation}

where $S_i^z$ is the spin on each site. The net magnetization vs external magnetic field plot for different $J$ values is depicted in \cref{fig:Results-HeisenberHamiltonianExMag-MSZ}. 
\begin{figure}
    \centering
    \includegraphics[scale = 0.58]{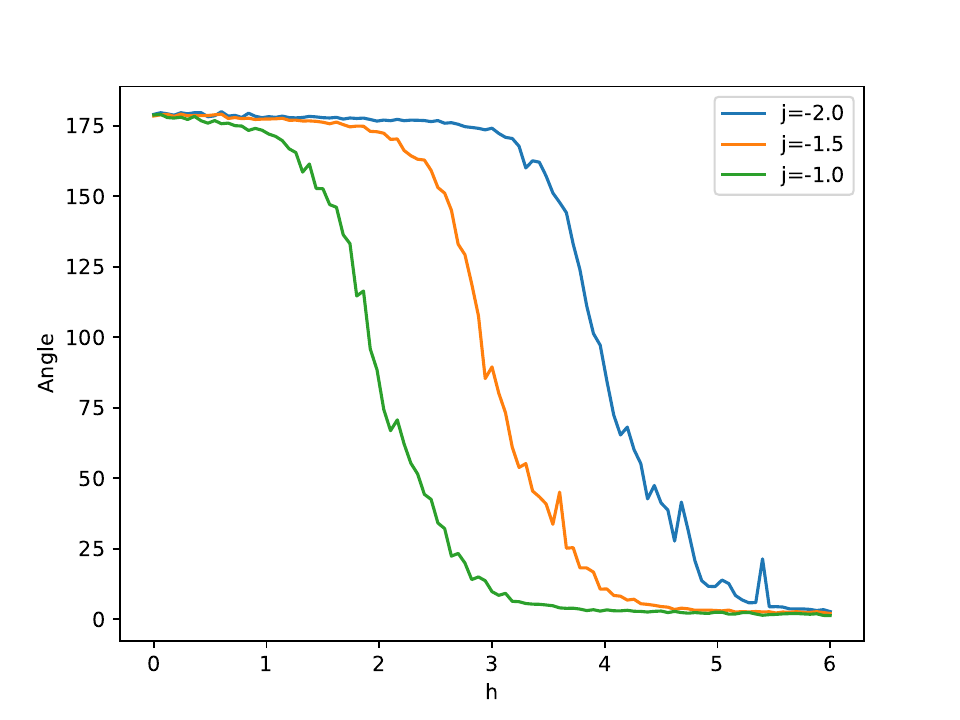}
    \caption{The angle between $\uparrow \& \downarrow$ spins vs external magnetic field graph for  $J \in [-1.0, -1.5, -2.0]$ with increasing magnetic field $h (\mu_B)$.}
    \label{fig:Results-HeisenberHamiltonianExMag-Angle}
\end{figure}
The net magnetization for the $J$ values in $\{-1.0, -1.5, -2.0\}$ and varying external magnetic field $h$ from $0.0$ to $6.0$ at the interval of $0.5$ is tabulated in \cref{tab:Results-HeisenberHamiltonianExMag-EnergyMSZ}, and the plot shows that with the increasing magnetic field, the net magnetization of the system starts increasing after a certain limit of the external magnetic field from its initial value of $0$ and becomes constant at certain $h$ at a value of $\sim 8$. This shows that when we increase the magnetic field, the spins try to keep aligned in anti-parallel order, but after a certain limit of $h$, the spin starts aligning in the direction of the external magnetic field. When we kept increasing the external magnetic field, the anti-parallel spins started aligning in the direction of the magnetic field, and all the spins became parallel to each other after a certain external field. With this, the starting antiferromagnetic ordering of the spins transformed into the ferromagnetic ordering, showing a magnetic phase transition. This magnetic phase transition is verified by plotting the angle between the anti-parallel spins with respect to the external magnetic field in \cref{fig:Results-HeisenberHamiltonianExMag-Angle}. From the plot, we can see that with no external field applied, $h = 0.0 \mu_B$, the angle between the anti-parallel spins is $\sim 180^{\circ}$, and with the increasing external magnetic field, the angle between the spin starts decreasing. After a certain $h$ value, the angle between the spins becomes $0^{\circ}$. This also justifies that initially, the spins were aligned in an anti-parallel direction, and with the increasing magnetic field, the spins aligned in the same direction.

\begin{figure}
    \centering
    \includegraphics[scale = 0.58]{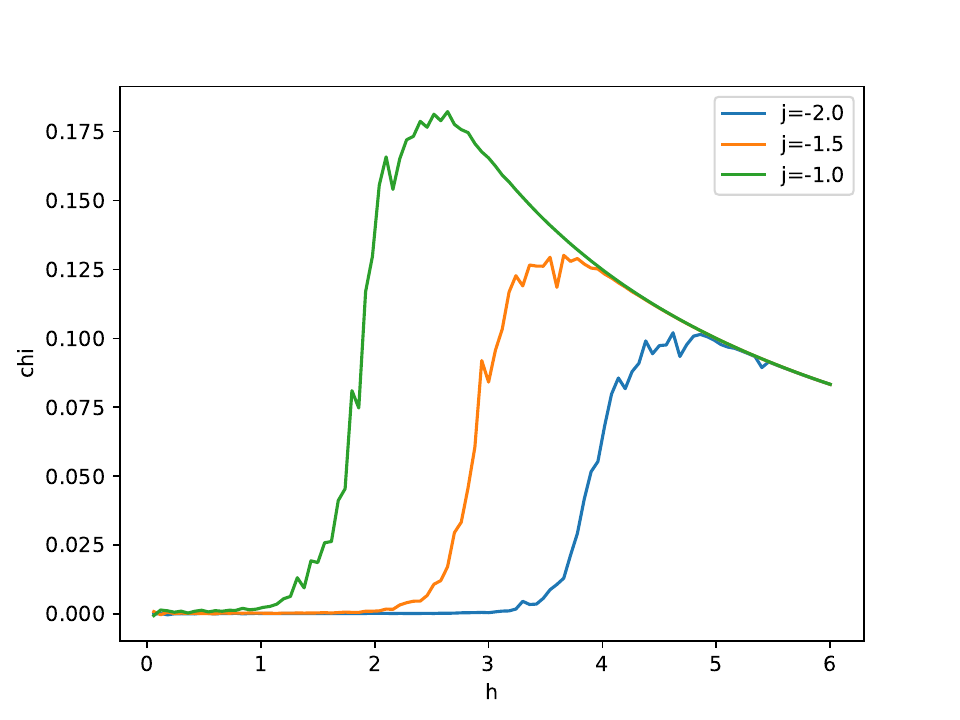}
    \caption{The magnetic susceptibility vs external magnetic field graph for  $J \in [-1.0, -1.5, -2.0]$ with increasing magnetic field $h (\mu_B)$.}
    \label{fig:Results-HeisenberHamiltonianExMag-chi}
\end{figure}

From the plot in \cref{fig:Results-HeisenberHamiltonianExMag-MSZ}, we also gathered that increasing $J$ value requires a higher external magnetic field $h$ to start the magnetic phase transition, which is consistent with the theory. As we know, $J$ is the coupling constant, and with increasing $J$-value, larger magnetic field is required to rotate the spins. The magnetic susceptibility vs external magnetic field plot is depicted in \cref{fig:Results-HeisenberHamiltonianExMag-chi}. The magnetic susceptibility of the quantum system is given as,
\begin{equation}
    \label{eq:Results-HeisenberHamiltonianExMag-chi}
    \chi_{\perp} = \frac{\partial M_{S^Z}}{\partial h}
\end{equation}

These net magnetization results show that the neural network was able to understand the effect of the magnetic field on the spins and net magnetization of the system. 

After applying the external magnetic field $\Vec{h} = h \hat{z}$ in the z-direction in Heisenberg Hamiltonian on a two-dimensional lattice with $M = 16$ sites occupied by $N = 16$ fermions, we found that the algorithm developed with the deep neural network was able to comprehend the effects of external magnetic field on the system by adjusting the neural network parameter in a way to accommodate those effects. From the results obtained from the simulation, we gathered that the algorithm is robust and can simulate any system for given external factors.

\section{\label{sec:conc}Conclusion}
This study aimed to assess the effectiveness of an Artificial Neural Network (ANN)-driven algorithm in simulating many-body fermionic systems. We presented the methodology for solving the fermionic system, including its mathematical formalism and simulation setup on one-dimensional and two-dimensional lattices. We constructed the quantum state using Pauli matrices and simulated the classical Heisenberg Hamiltonian, examining its response to an external magnetic field.

The constructed ANN architecture for simulating the fermionic system was integrated into the algorithm. Subsequently, a sampling process using Markov Chain Monte Carlo importance sampling, particularly employing the exchange sampler to preserve antiferromagnetic ordering, was formulated. The neural network optimization was performed using an adaptive momentum optimizer.

In the results section, we demonstrated the algorithm's efficacy. For a one-dimensional lattice with ten sites occupied by ten fermions, we observed efficient convergence to the ground state energy, reaching approximately $-9.0 ~\text{eV}$ from an initial state near $-5.5 ~\text{eV}$. The per-site ground state energy exhibited high accuracy, with our algorithm yielding $-0.9028 ~\text{eV}$ compared to the exact diagonalization result of $-0.9031 ~\text{eV}$. Moreover, the simulation of a two-dimensional lattice confirmed these findings and revealed the system's behaviour under varying external magnetic fields.

Our study concludes that the ANN-driven algorithm robustly simulates fermionic systems, accurately computing their properties and efficiently adapting to external influences. Notably, it outperforms traditional Monte Carlo and exact diagonalization methods, offering comparable or superior accuracy with reduced computational costs. Moreover, it effectively handles frustrated systems without encountering the sign problem.

In summary, our research underscores the potential of ANN-driven algorithms in quantum simulations, offering insights into complex many-body systems and paving the way for further advancements in this field.
\begin{acknowledgments}
Financial support for the DST India INSPIRE fellowship through grant number IF171000, SERB India through grant number CRG/2021/005320, and the use of the high-performance computing facility at IISER Bhopal are gratefully acknowledged.
\end{acknowledgments}
\bibliography{library}

%
\end{document}